\documentclass[aps,pra,reprint,superscriptaddress,showpacs]{revtex4-2}
\usepackage{graphicx}
\usepackage{dcolumn}
\usepackage{bm}
\usepackage{subfigure}
\usepackage{algorithm}
\usepackage{algpseudocode}
\usepackage{appendix}
\usepackage{float}
\usepackage[braket, qm]{qcircuit}
\usepackage{mathrsfs}
\usepackage{longtable}
\usepackage{amsfonts}
\usepackage{amstext}
\usepackage[usenames]{xcolor}
\usepackage{epsfig}
\usepackage{epstopdf}
\usepackage{mathrsfs}
\usepackage{bm}
\usepackage{amsmath}
\usepackage{amssymb,amsthm}
\usepackage{tikz}
\usepackage{hyperref}
\usepackage{comment}

\hypersetup{colorlinks=true, citecolor=blue, urlcolor=blue, linkcolor=blue}

\begin{document}
\newtheorem{theorem}{Theorem}
\newtheorem{thm}{Theorem}
\newtheorem{cor}[thm]{Corollary}
\newtheorem{lem}[thm]{Lemma}
\newtheorem{prop}[thm]{Proposition}
\newtheorem{propp}{Proposition}
\newtheorem*{pf}{Proof}
\newtheorem{definition}{Definition}
\newtheorem{ex}{Example}
\newtheorem{remark}{Remark}
\newtheorem*{thmm}{Theorem}
\newtheorem*{corr}{Corollary}

\title{Decoding Quantum Search Advantage: The Critical Role of State Properties in Random Walks}

\author{Si-Qi Zhou}
\affiliation{School of Computer Science and Engineering, Sun Yat-sen University, Guangzhou 510006, China}
\affiliation{School of Mathematical Sciences, 
MOE-LSC, Shanghai Jiao Tong University, Shanghai, 200240, China}
\author{Jin-Min Liang}
\email{jinmin.liang@btbu.edu.cn}
\affiliation{School of Mathematics and Statistics, Beijing Technology and Business University, Beijing 100048, China}
\author{Ziheng Ding}
\affiliation{School of Mathematical Sciences, MOE-LSC, Shanghai Jiao Tong University, Shanghai, 200240, China}
\affiliation{Shanghai Seres Information Technology Co., Ltd, Shanghai, 200040, China} \affiliation{Shenzhen Institute for Quantum Science and Engineering, Southern University of Science and Technology, Shenzhen, 518055, China}
\author{Zhihua Chen}
\email{chenzhihua77@sina.com}
\affiliation{School of Science, Jimei University, Xiamen 361021, China}
\author{Shao-Ming Fei}
\email{feishm@cnu.edu.cn}
\affiliation{School of Mathematical Sciences, Capital Normal
University, Beijing 100048, China}
\affiliation{Max Planck Institute for Mathematics in the Sciences - 04103 Leipzig, Germany}
\author{Zhihao Ma}
\email{mazhihao@sjtu.edu.cn}
\affiliation{School of Mathematical Sciences, MOE-LSC, Shanghai Jiao Tong University, Shanghai, 200240, China}
\affiliation{Shanghai Seres Information Technology Co., Ltd, Shanghai, 200040, China}
\affiliation{Shenzhen Institute for Quantum Science and Engineering, Southern University of Science and Technology, Shenzhen, 518055, China}

\begin{abstract}
Quantum algorithms have demonstrated provable speedups over classical counterparts, yet establishing a comprehensive theoretical framework to understand the quantum advantage remains a core challenge. In this work, we decode the quantum search advantage by investigating the critical role of quantum state properties in random-walk-based algorithms. We propose three distinct variants of quantum random-walk search algorithms and derive exact analytical expressions for their success probabilities. These probabilities are fundamentally determined by specific initial state properties: the coherence fraction governs the first algorithm's performance, while entanglement and coherence dominate the outcomes of the second and third algorithms, respectively. We show that increased coherence fraction enhances success probability, but greater entanglement and coherence reduce it in the latter two cases. These findings reveal fundamental insights into harnessing quantum properties for advantage and guide algorithm design. Our searches achieve Grover-like speedups and show significant potential for quantum-enhanced machine learning.
\end{abstract}

\maketitle

\noindent\textbf{Keywords:} quantum random-walk search algorithm, quantum coherence fraction, quantum entanglement, quantum coherence

\noindent\textbf{Introduction --}
Quantum computing has emerged as a revolutionary computational paradigm that promises to solve certain problems exponentially faster than classical computers~\cite{bennett2000quantum,365700,grover1996fast, PhysRevLett.79.325,feynman2018simulating, RevModPhys.86.153,PhysRevLett.103.150502,ladd2010quantum,harrow2017quantum, neill2018blueprint, yung2019quantum,biamonte2017quantum,tang2019quantum,yu2021quantum, west2023towards}. Notable examples of these algorithms include Shor’s factoring algorithm~\cite{365700}, Grover’s search algorithm~\cite{grover1996fast, PhysRevLett.79.325}, simulations of various quantum systems~\cite{feynman2018simulating, RevModPhys.86.153}, HHL's algorithm for linear equation systems~\cite{PhysRevLett.103.150502}, and so on~\cite{ladd2010quantum,harrow2017quantum, neill2018blueprint}. While remarkable theoretical advances have been made in developing quantum algorithms with provable speedups, a comprehensive understanding of the fundamental origins of these quantum advantages remains elusive - which is critically important for designing practical quantum algorithms with real-world applications\cite{arute2019quantum, zhong2020quantum,ekert1998quantum,PhysRevA.100.012349,ahnefeld2022coherence}. We delve into the quantumnesses of quantum states that affect the efficiency of quantum computation, focusing on how such quantumnesses can be leveraged to optimize quantum algorithms. This knowledge gap becomes particularly pressing in NISQ era~\cite{Preskill2018quantumcomputingin,doi:10.1126/sciadv.adr5002}, where practical implementations face significant constraints from limited qubit counts and substantial noise, making it essential to identify and harness the specific quantum resources that can deliver robust computational advantages.

It is generally believed that quantum entanglement~\cite{PhysRevLett.78.2275,RevModPhys.81.865} plays a key role in enabling a variety of quantum computational tasks~\cite{PhysRevLett.87.077902,PhysRevLett.110.060504,PhysRevX.5.041008}. In parallel, the quantum coherence~\cite{PhysRevLett.113.140401}, as one of the essential quantum properties originated from the superposition principle of quantum states, plays an important role in quantum information processing~\cite{PhysRevLett.116.080402,PhysRevLett.116.160407,PhysRevLett.117.030401,PhysRevLett.120.230504,hu2018quantum}. Besides, the fully entangled fraction~\cite{PhysRevA.54.3824} and the quantum coherence fraction~\cite{PhysRevA.100.032324} also characterize the important properties of quantum states, serving as the effective indicators~\cite{PhysRevA.62.012311,PhysRevA.66.012301,PhysRevLett.127.080502,PhysRevA.93.032326}. Quantum entanglement, coherence and coherence fraction are intrinsic properties of quantum states and have been shown to play crucial roles in enhancing the performance and efficiency of quantum algorithms~\cite{ekert1998quantum,PhysRevA.65.062312,PhysRevA.93.012111,etde_21590818,PhysRevA.95.032307,PhysRevA.100.012349,PhysRevA.99.032320,PhysRevLett.129.120501,PhysRevA.110.062429}.

Among various quantum computational approaches, quantum random walks have attracted considerable attention due to their intrinsic parallelism and potential for implementing practical quantum algorithms under current technological constraints. Recent investigations on quantum walks~\cite{PhysRevA.48.1687} have revealed that, unlike their classical counterparts, quantum walks exhibit unique dynamical properties, leveraging quantum coherence and interference to achieve computational advantages. Over the past three decades, quantum walks have emerged as a powerful framework with profound implications across multiple domains, including quantum computation, quantum information theory, and fundamental physics~\cite{PhysRevA.67.052307,science.1260364,PhysRevLett.119.220503,PhysRevLett.121.070402}. Notably, quantum walks provide an intuitive yet highly effective approach for designing novel and efficient quantum algorithms, reinforcing a fundamental role in advancing quantum computing~\cite{PhysRevA.67.052307,PhysRevA.70.022314,1360855568517402880,PhysRevA.104.022216,PhysRevA.111.042608}. Furthermore, quantum walks have emerged as a pivotal tool in machine learning, a key domain of artificial intelligence, underscoring the transformative potential of quantum computing in advancing modern data processing techniques~\cite{PhysRevLett.110.220501,9321551,PhysRevA.107.042405,flamini2024towards}.
However, the precise quantum resources enabling this advantage - and how they should be optimally utilized - remain poorly understood.

In 2003, Shenvi, Kempe, and Whaley (SKW)~\cite{PhysRevA.67.052307} introduced a quantum search algorithm based on discrete-time quantum walks on an $n$-dimensional hypercube. Unlike Grover’s algorithm, which amplifies amplitudes globally, the SKW algorithm employs interference through localized transitions, offering a graph-structured and physically intuitive search mechanism. While both algorithms achieve the same query complexity, the SKW model provides a conceptually distinct framework that naturally integrates with graph-based computation and is more amenable to generalization. Its structured evolution also suggests advantages in robustness to noise and experimental feasibility. These properties make SKW a compelling subject for further investigation in the development of quantum walk-based search algorithms. Various enhancements and experimental implementations of the SKW algorithm have been proposed~\cite{PhysRevA.78.012310,PhysRevA.79.012325,PhysRevA.81.022308}.

\begin{figure}[h]
\centering
\includegraphics[width=2.5in]{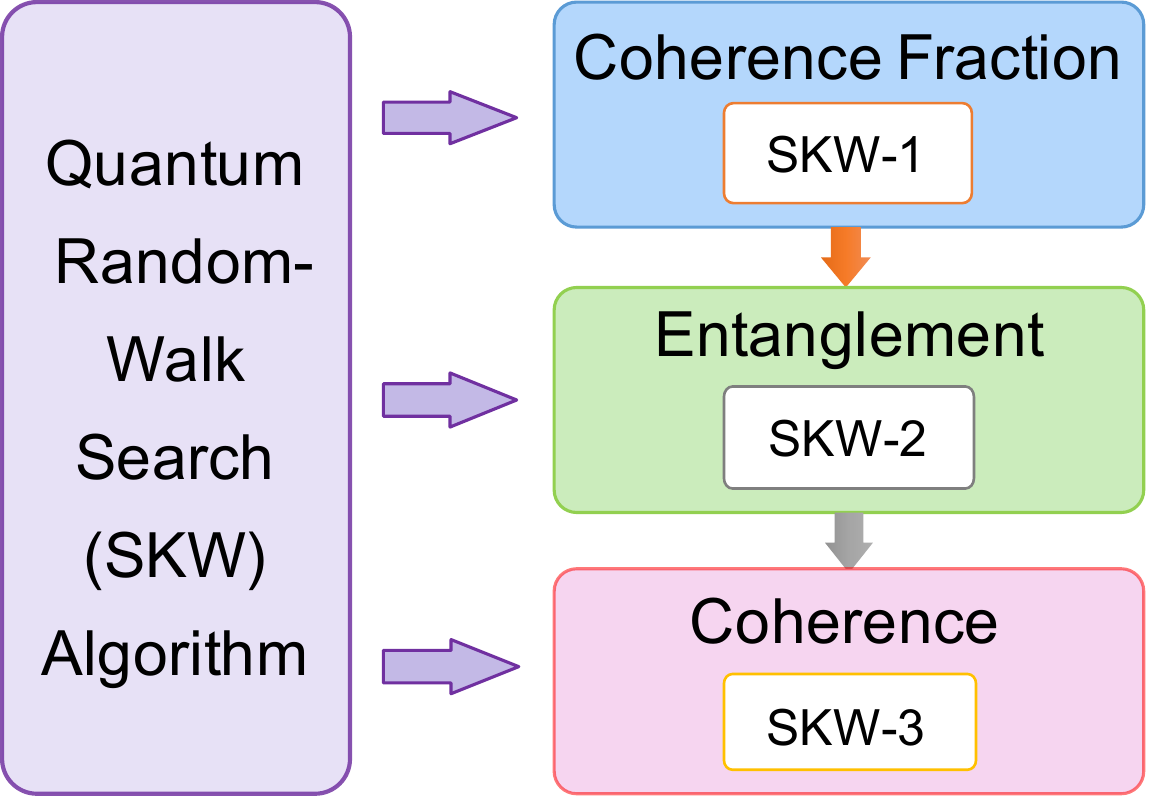}
\caption{\textbf{Diagrammatic sketch of the quantum properties of the initial state in SKW algorithm.} Three modified versions of the SKW algorithm give connections between the algorithm's success probabilities and the coherence fraction, entanglement and coherence, respectively.}
\label{fig:SKW-0}
\end{figure}

In this work, we investigate the SKW algorithm by revisiting its success probability, with a particular focus on the quantum properties of the initial state. We present three modified versions of the SKW algorithm and explore the relationships between the success probabilities of these algorithms and the coherence fraction, entanglement, and coherence, as illustrated in Fig.~\ref{fig:SKW-0}. We show that a higher coherence fraction leads to increased success probability for the first algorithm, while higher entanglement and coherence result in decreased success probabilities for the second and third algorithms. Our findings provide practical insights for designing quantum algorithms with provable advantages.

\noindent\textbf{Quantum random-walk search algorithm --}\label{algo0}
We first briefly recall the SKW algorithm.
The SKW algorithm~\cite{PhysRevA.67.052307} comprises two components: a quantum component and a classical protocol that embeds it. The quantum part involves a perturbed Grover walk on a hypercube, starting from an equally weighted superposition of initial states and iteration for a specified number of steps. This is followed by a measurement of the output state to identify the marked vertex. The perturbation of the Grover coin is introduced by the oracle, which imparts position dependence to the coin operator. The SKW algorithm has been shown to require $O(\sqrt{N})$ number of oracle queries to find the marked element, where $N$ is the size of the search space. The search space is defined as the set of all $n$-bit binary strings, $\vec{x}=\{0, 1\}^n$. Consider a function $f(\vec{x})$, where $f(\vec{x})=1$ for exactly the input $\vec{x}_{\mathrm{tg}}$. The goal of this search algorithm is to find $\vec{x}_{\mathrm{tg}}$. By mapping the $n$-bit binary string to nodes on the hypercube, this search problem becomes equivalent to searching for a single marked node among $N=2^n$ nodes on the $ n$-dimensional cube.

The SKW quantum walk takes place on the product Hilbert space $\mathcal{H}=\mathcal{H}^{C_{n}}\otimes \mathcal{H}^{V_{n}}$, where $\mathcal{H}^{V_{n}}$ is the $N=2^{n}$-dimensional Hilbert space representing the vertices, and $\mathcal{H}^{C_{n}}$ is the $n$-dimensional space associated with the quantum coin. Each state in $\mathcal{H}$ can be described by a bit string $\vec{x}$, which specifies the position on the hypercube, and a direction $d$, which specifies the state of the coin. The shift operator $\mathcal{S}$ maps a state $|d, \vec{x}\rangle$ onto the state $|d, \vec{x}\oplus \vec{e}_{d} \rangle$, where $\vec{e}_{d}$ is the $d$th basis vector on the hypercube, corresponding to the edges originated from the given vertex. $\mathcal{S}$ can be written explicitly as $\mathcal{S}=\sum_{d, \vec{x}}|d, \vec{x} \oplus \vec{e}_{d}\rangle\langle d, \vec{x}|$. If the target vertex marked is denoted by $\vec{x}_{\mathrm{tg}}$, the perturbed coin operator can be written as $\mathcal{C}=\mathcal{C}_{0} \otimes \mathbf{I}+(\mathcal{C}_{1}-\mathcal{C}_{0}) \otimes|\vec{x}_{\mathrm{tg}}\rangle\langle \vec{x}_{\mathrm{tg}}|$. $\mathcal{C}_{0}$ is usually chosen to be the Grover operator $\mathcal{G}$ and $\mathcal{C}_{1}$ is chosen to be $-\mathbf{I}$. The SKW algorithm initializes the quantum computer to the equal superposition over all states: $|S^c\rangle \otimes |S^s\rangle$, where $|S^c\rangle=\frac{1}{\sqrt{n}}\sum_{d=1}^{n}|d\rangle$ and $|S^s\rangle=\frac{1}{\sqrt{N}}\sum_{x=0}^{N-1}|\vec{x}\rangle$. Note that $|S^s\rangle$ is the equal superposition state that can be prepared efficiently on the node space by applying $n$-bit Hadamard operations to the $|\vec{0}^n\rangle$ state, similarly for $|S^c\rangle$. 

The measurement outcomes will be the marked state with probability $P=\frac{1}{2}-O(\frac{1}{\sqrt{N}})$. By repeating the algorithm a constant number of times, the marked state will be determined with an arbitrarily small degree of error. Various optimizations and improvements of the SKW algorithm have also been proposed. In particular, Potoček et al.~\cite{PhysRevA.79.012325} optimized the SKW algorithm with a significant increase in the success probability and an improvement on query complexity so that the theoretical limit of a search algorithm succeeding with probability close to one is attained. Subsequently, an experiment has been conducted to demonstrate the $1$ out of $4$ case of the SKW algorithm, showcasing its superiority over classical algorithms~\cite{PhysRevA.81.022308}. 

\noindent\textbf{Modified quantum random-walk search algorithms --}\label{algo}
We analyze below the SKW algorithm by investigating how the quantum properties of the initial state influence the success probability. We propose three modified versions of the algorithm that establish links between success probability and the coherence fraction, entanglement and coherence, respectively.

\textit{Coherence fraction in SKW-1 algorithm.}
The coherence fraction of a state $\rho$ is defined by the Uhlmann's fidelity between states $\rho$ and $|\eta\rangle $~\cite{PhysRevA.100.032324},
\begin{equation}\label{f_{c}}
f_{c}(\rho):= F(|\eta\rangle, \rho)=\langle \eta | \rho |\eta\rangle,  
\end{equation}
where $|\eta\rangle=\sum_{x=0}^{N-1}|x\rangle/\sqrt{N}$ is the equal superposition state (the maximal coherent state), the Uhlmann fidelity between two general states $\rho$ and $\sigma$ is given by $F(\rho, \sigma) \equiv [\operatorname{Tr}(\sqrt{\rho} \sigma \sqrt{\rho})^{1 / 2}]^2$. The coherence fraction quantifies the closeness of a quantum state to the maximal coherent state, in analogy to the entangled fraction.

\begin{figure}[h]
\centering
\includegraphics[width=3in]{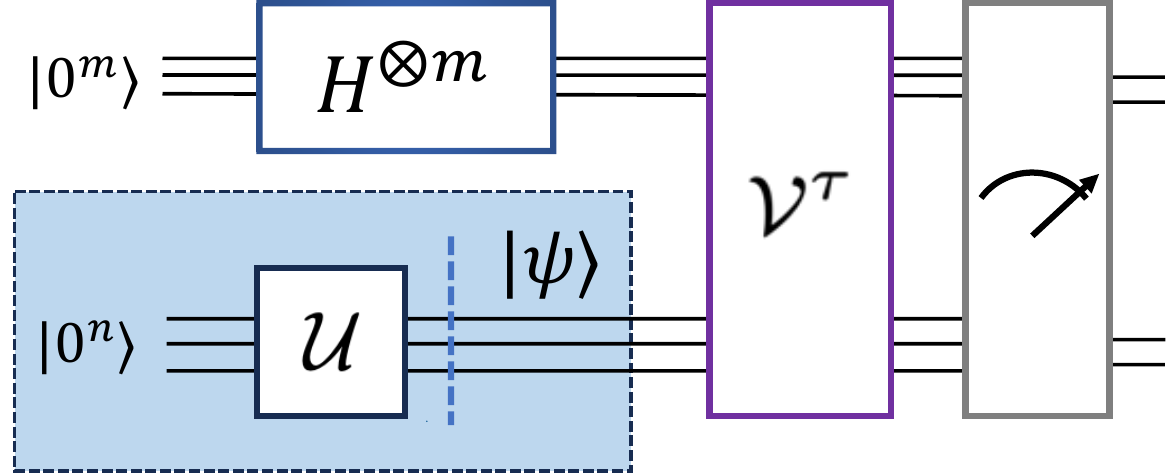}
\caption{\textbf{Quantum circuit for SKW-1 algorithm.} Applying the Hadamard operation $H^{\otimes m}$ ($m^2=n$) to the input state $|0^m\rangle$ on the direction space to obtain $|S^c\rangle=\frac{1}{\sqrt{n}}\sum_{d=1}^{n}|d\rangle$. While an arbitrary unitary quantum gate $\mathcal{U}$ is applied to the input state $|\vec{0^{n}}\rangle$ of the register. The initial state is $|\psi\rangle=\mathcal{U}|\vec{0^{n}}\rangle=\sum_{x=0}^{N-1}a_x|\vec{x}\rangle$, where $a_x$ is the amplitude of $|\vec{x}\rangle$. The resulting state is $|S^c\rangle\otimes|\psi\rangle$. Subsequently, the perturbed evolution operator $\mathcal{V}=\mathcal{S}\mathcal{C}$ is applied $\tau$ times, and the final state is then measured.}
\label{fig:SKW-1}
\vspace{-5mm}
\end{figure}

We propose a modified quantum random-walk search algorithm (SKW-1) detailed in Algo.~\ref{alg:SKW-1}, with its corresponding circuit depicted in Fig.~\ref{fig:SKW-1}. We are interested in how the initial state impacts the algorithm's search performance and which intrinsic quantum properties contribute to the SKW-1 algorithm. In contrast to the Hadamard gate used in the node space of SKW algorithm, the SKW-1 algorithm implements an arbitrary unitary gate $\mathcal{U}$ before the perturbed evolution operator $\mathcal{V}$ to obtain the pure initial state $|\psi\rangle=\sum_{x=0}^{N-1}a_x|\vec{x}\rangle$ with amplitude $a_x$. The selection of an arbitrary initial state $|\psi\rangle$ indicates that the weights assigned to the vertices in the node space are not necessarily equal, but can instead be chosen freely, thus allowing for a more generalized and adaptable framework for the system. Additionally, it provides a direct generalization to the case of the initial mixed state $\rho$. 

\begin{algorithm}[H]
\caption{SKW-1 algorithm}
\vspace{0.7em}
\begin{itemize}
\item [1.] Initialize the quantum state $|\Phi\rangle = |S^c\rangle \otimes |\psi\rangle$, where $|S^c\rangle= H^{\otimes m}|0^m\rangle=\frac{1}{\sqrt{n}}\sum_{d=1}^{n}|d\rangle \quad (m^2 = n)$, $|\psi\rangle= \mathcal{U}|0^n\rangle=\sum_{x=0}^{N-1}a_x|\vec{x}\rangle$, $H$ is the Hadamard gate, and $\mathcal{U}$ is an arbitrary unitary quantum gate.

\item [2.] Apply the perturbed evolution operator $\mathcal{V}=\mathcal{S}\mathcal{C}$ about $\tau$ times.
$\mathcal{S}=\sum_{d, \vec{x}}|d, \vec{x} \oplus \vec{e}_{d}\rangle\langle d, \vec{x}|$ is the shift operator. The perturbed coin operator for the target vertex $|\vec{x}_{\mathrm{tg}}\rangle$ is $\mathcal{C}=\mathcal{C}_{0} \otimes \mathbf{I}+(\mathcal{C}_{1}-\mathcal{C}_{0}) \otimes|\vec{x}_{\mathrm{tg}}\rangle\langle \vec{x}_{\mathrm{tg}}|$, where $\mathcal{C}_{0}$ is usually chosen to be the Grover operator $\mathcal{G}$ and $\mathcal{C}_{1}$ is chosen to be $-\mathbf{I}$.

\item [3.] Measure the state.
\end{itemize}
\label{alg:SKW-1}
\end{algorithm}

Now, we present our main result related to the SKW-1 algorithm, see proof in \cite[Sec.~I]{SuppMaterial}.

\begin{theorem}\label{thm1}
For any given initial state $|\psi\rangle$ in the node space, applying $O(\sqrt{N})$ iterations, the average success probability of the SKW-1 algorithm over all $N$ possible target states is upper bounded by
\begin{equation}\label{max-1}
\begin{aligned}
P_{\text{max-1}}(|\psi\rangle)
=\frac{1}{2}f_{c}(|\psi\rangle)+O\left(\frac{1}{\sqrt{N}}\right),
\end{aligned}
\end{equation}
where $f_{c}(|\psi\rangle)$ is the coherence fraction of the initial state defined by the fidelity $F(|\eta\rangle, |\psi\rangle)$ between $|\psi\rangle$ and the equal superposition state $|\eta\rangle=\sum_{x=0}^{N-1}|\vec{x}\rangle/\sqrt{N}$.
\end{theorem}

Eq.(\ref{max-1}) implies that the average success probability $P_{\text{max-1}}(|\psi\rangle)$ is determined exclusively by the coherence fraction of the initial state, $f_{c}(|\psi\rangle)=|\langle \eta | \psi \rangle|^2$.

Neglecting the term $O(1/\sqrt{N})$~\cite{PhysRevA.65.062312}, we have $P_{\text{max-1}}(|\psi\rangle)\in[0, 1/2]$ as $f_{c}(|\psi\rangle)\in[0,1]$. The upper bound of $P_{\text{max-1}}(|\psi\rangle)$ is attained when the initial state is $|\psi\rangle=|\eta\rangle=|+\rangle^{\otimes n}$, which corresponds to the original SKW algorithm. This indicates that the equal superposition state gives rise to the highest success probability. Our findings explain why the Hadamard gate is typically used to generate the equal superposition state, rather than an arbitrary quantum unitary gate. Note that our previous analysis focused on pure states. Similar results are given in \cite[Sec.~II]{SuppMaterial} for mixed initial states. We generalize our results to the case of optimized quantum random-walk search (OSKW) algorithm in \cite[Sec.~III]{SuppMaterial}.

The SKW-1 algorithm reveals that the success probability is determined exclusively by the coherence fraction of the initial state, rather than the entanglement or coherence of the initial state. To explore the full potential of quantum computing, it is vital to understand which quantum properties offer computational efficiency. Various quantum resources, such as entanglement and coherence, have played significant roles in quantum information processing. We further focus on the roles played by the entanglement and coherence in other modified SKW algorithms.

\textit{Entanglement in SKW-2 algorithm.}
Here, we consider the entanglement of the initial state. As in SKW-1 algorithm, the initial node space state in SKW-2 algorithm is prepared as an arbitrary pure state of the form $|\psi\rangle = \sum_{x=0}^{N-1} a_x |\vec{x}\rangle$, where the complex amplitudes $a_x$ are freely specifiable. We show that the maximal success probability $P_{\text{max-2}}$ over all possible local unitary operations in the initialization step is related to the entanglement present in the initial register state $|\psi\rangle$. Here we adopt the Groverian entanglement measure $E_g(|\psi\rangle)$ of a state $|\psi\rangle$~\cite{PhysRevA.65.062312},
\begin{equation}\label{ent}
    E_g(|\psi\rangle)\equiv \min _{\sigma \in \mathsf{S}} \sqrt{1-F(\sigma, |\psi\rangle)},
\end{equation}
where $\mathsf{S}$ is the set of all separable states.

\begin{figure}[t]
\centering
\includegraphics[width=3in]{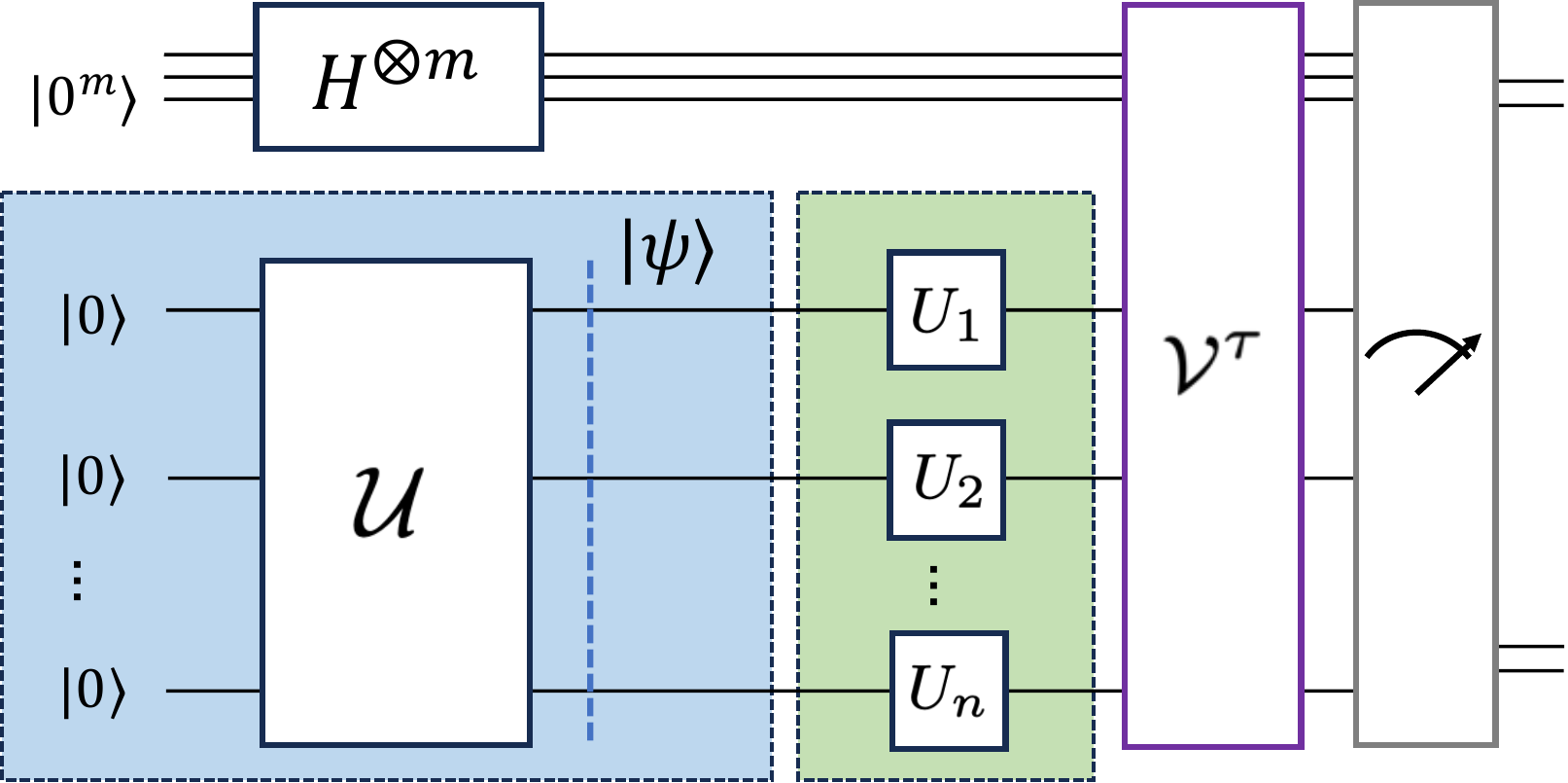}
\caption{\textbf{Quantum circuit for SKW-2 algorithm.} Applying the Hadamard operation $H^{\otimes m}$ ($m^2=n$) to the input state $|0^m\rangle$ on the direction space to obtained $|S^c\rangle=\frac{1}{\sqrt{n}}\sum_{d=1}^{n}|d\rangle$. While an arbitrary unitary quantum gate $\mathcal{U}$ is applied to the input state $|\vec{0^{n}}\rangle$ of the register. The initial state is $|\psi\rangle=\mathcal{U}|\vec{0^{n}}\rangle=\sum_{x=0}^{N-1}a_x|\vec{x}\rangle$, where $a_x$ is the amplitude of $|\vec{x}\rangle$. The resulting state is $|S^c\rangle|\psi\rangle$. Perform a product of arbitrary local operations $U_{1}\otimes U_{2}\otimes \cdots \otimes U_{n}$ on the register, where $U_{j}$ is an arbitrary local unitary gate acting on the $j$th qubit. Subsequently, the perturbed evolution operator $\mathcal{V}=\mathcal{S}\mathcal{C}$ is applied $\tau$ times, and the final state is measured on the computational basis.}
\label{fig:SKW-2}
\vspace{-7mm}
\end{figure}

Consider $n$ parties sharing an $n$-qubit pure quantum state $|\psi\rangle$. For simplicity, we initially assume that each party owns one qubit. This modified SKW algorithm (SKW-2) is summarized in Algo.~\ref{alg:SKW-2}, with its corresponding circuit given in Fig.~\ref{fig:SKW-2}.

\begin{algorithm}[H]
\caption{SKW-2 algorithm}
\vspace{0.7em}
\begin{itemize}
\item [1.] Initialize the quantum state $|\Phi\rangle = |S^c\rangle \otimes |\psi\rangle$, where $|S^c\rangle= H^{\otimes m}|0^m\rangle=\frac{1}{\sqrt{n}}\sum_{d=1}^{n}|d\rangle \quad (m^2 = n)$, $|\psi\rangle= \mathcal{U}|0^n\rangle=\sum_{x=0}^{N-1}a_x|\vec{x}\rangle$, $H$ is the Hadamard gate, and $\mathcal{U}$ is an arbitrary unitary quantum gate.

\item [2.] Apply a product of arbitrary local operations $U_{1}\otimes U_{2}\otimes \cdots \otimes U_{n}$ on the state $|\psi\rangle$ in node space, where $U_{j}$ is an arbitrary unitary gate acting on the $j$th qubit. The resulting state is $|S^c\rangle\otimes(U_{1}\otimes U_{2}\otimes \cdots \otimes U_{n})|\psi\rangle$.

\item [3.] Repeat $\tau$ times of the perturbed evolution operator $\mathcal{V}=\mathcal{S}\mathcal{C}$. 

\item [4.] Measure the state of the register. 
\end{itemize}
\label{alg:SKW-2}
\end{algorithm}

We have the following result for the SKW-2 algorithm, and the full proof is included in \cite[Sec.~IV]{SuppMaterial}.

\begin{theorem}\label{thm2}
For any given initial state $|\psi\rangle$ in the node space, the maximal success probability of the SKW-2 algorithm, averaged over all $N$ possible target states, after applying $O(\sqrt{N})$ iterations, is given by
\begin{equation}\label{max-2}
P_{\text{max-2}}=\frac{1-E_g^2(|\psi\rangle)}{2},
\end{equation}
where $E_g(\psi)$ is the Groverian entanglement measure~\cite{PhysRevA.65.062312} of a state $|\psi\rangle$ defined in Eq.(\ref{ent}).
\end{theorem}

Since $0 \leqslant E_g(|\psi\rangle) \leqslant 1$, it follows that $0 \leqslant P_{\max } \leqslant \frac{1}{2}$. The theorem shows that $P_{max}$ depends on the Groverian entanglement $E_g(|\psi\rangle)$ of the initial register state $|\psi\rangle$. Similar results can be obtained for the optimized algorithm OSKW. The corresponding success probability is $P_{\text{max-2}}^{\text{opt}}=1-E_g^2(|\psi\rangle)$, which is also uniquely determined by the Groverian entanglement $E_g(|\psi\rangle)$.

Eq. (\ref{max-2}) demonstrates that the SKW-2 algorithm's success probability decreases with increasing initial state entanglement. While entanglement is typically a valuable quantum resource, it detrimentally impacts the SKW-2 algorithm. This finding highlights a paradox that excessive initial entanglement may harm performance, indicating a need for careful balancing in the design of quantum algorithms.

\textit{Coherence in SKW-3 algorithm.}
To investigate the performance of the coherence of the initial state, we consider the maximal probability of success $P_{\text{max-3}}$ over all possible local unitary operations chosen from the three single-qubit Pauli gates (X, Y, and Z) before the perturbed evolution step. The maximal probability of success $P_{\text{max-3}}$ is related to the coherence present in the initial register state $|\psi\rangle$. We adopt the coherence measure based on fidelity~\cite{liu2017new},
\begin{equation}\label{coh}
    C_f(|\psi\rangle)\equiv \min _{\delta \in \mathcal{I}} \sqrt{1-F(\delta, |\psi\rangle)},
\end{equation}
where $\mathcal{I}$ is the set of all incoherent states, that is, the diagonal density matrices in the given basis.

\begin{figure}[h]
\centering
\includegraphics[width=2.7in]{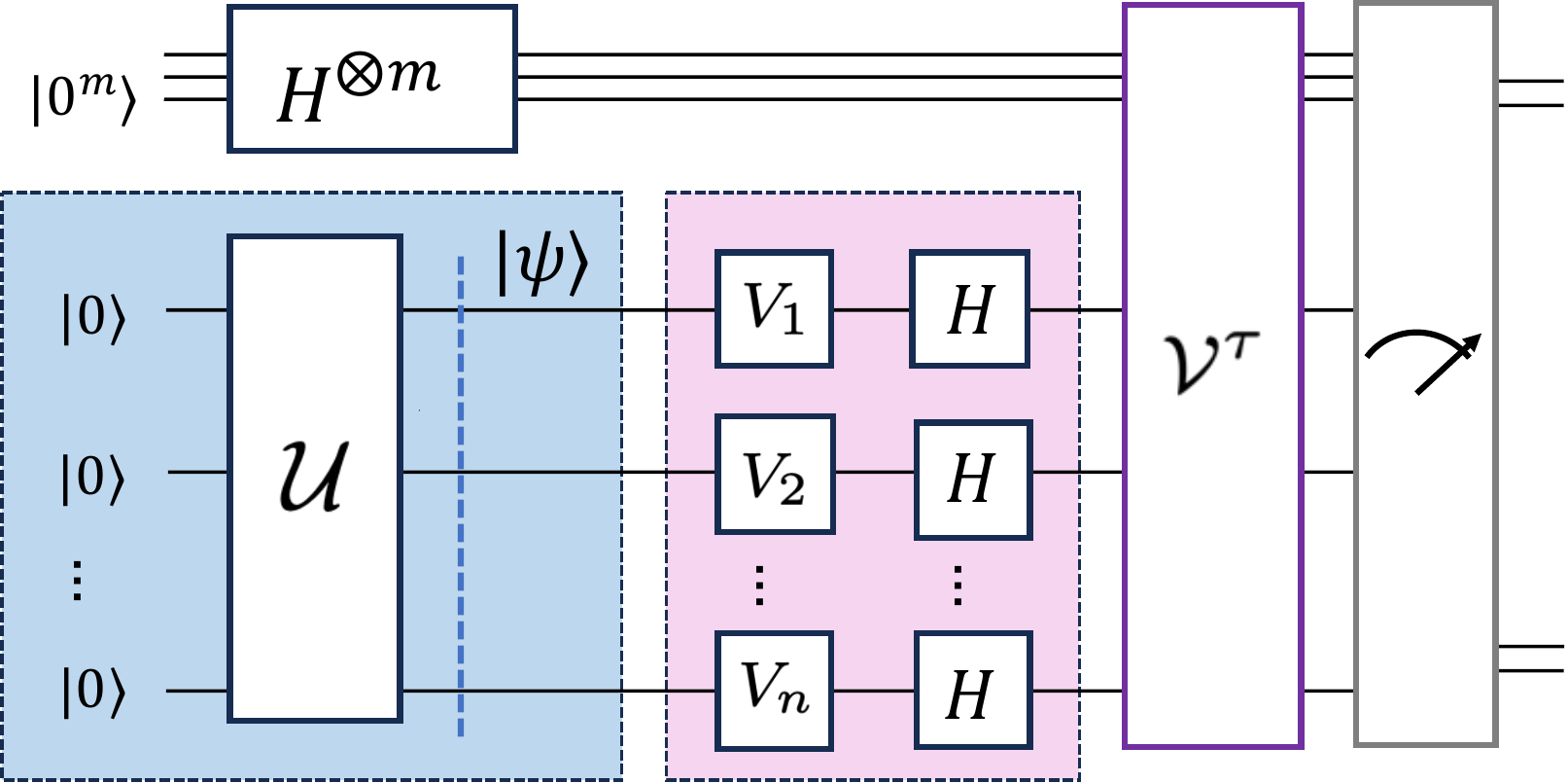}
\caption{\textbf{Quantum circuit for SKW-3 algorithm.} Applying the Hadamard operation $H^{\otimes m}$ ($m^2=n$) to the input state $|0^m\rangle$ on the direction space to obtained $|S^c\rangle=\frac{1}{\sqrt{n}}\sum_{d=1}^{n}|d\rangle$. While an arbitrary unitary quantum gate $\mathcal{U}$ is applied to the input state $|\vec{0^{n}}\rangle$ of the register. The initial state is $|\psi\rangle=\mathcal{U}|\vec{0^{n}}\rangle=\sum_{x=0}^{N-1}a_x|\vec{x}\rangle$, where $a_x$ is the amplitude of $|\vec{x}\rangle$. The resulting state is $|S^c\rangle|\psi\rangle$. Apply a product of arbitrary local operations $V_{1}\otimes V_{2}\otimes \cdots \otimes V_{n}$ on the initial state $|\psi\rangle$, where $V_{j}$ is chosen from the three single-qubit Pauli gates (X, Y, and Z) acting on the $j$th qubit. Then perform a Hadamard gate $H$ to each qubit in the register. Subsequently, the perturbed evolution operator $\mathcal{V}=\mathcal{S}\mathcal{C}$ is applied $\tau$ times, and the final state is measured on the computational basis.}
\label{fig:SKW-3}
\vspace{-7mm}
\end{figure}

In SKW-3 algorithm, the initial state of the node space is likewise prepared as an arbitrary pure state of the form $|\psi\rangle = \sum_{x=0}^{N-1} a_x |\vec{x}\rangle$, with freely specifiable complex amplitudes $a_x$. This setup maintains consistency with the initialization used in SKW-1 and SKW-2 algorithms. Consider $n$ parties sharing an $n$-qubit pure state $|\psi\rangle$. Each party owns one qubit. The modified ad hoc algorithm (SKW-3) is summarized in Algo.~\ref{alg:SKW-3}. Its corresponding circuit is presented in Fig.~\ref{fig:SKW-3}.

We have the following conclusion for the SKW-3 algorithm, see the proof in \cite[Sec.~IV]{SuppMaterial}.

\begin{algorithm}[H]
\caption{SKW-3 algorithm}
\vspace{0.7em}
\begin{itemize}
\item [1.] Initialize the quantum state $|\Phi\rangle = |S^c\rangle \otimes |\psi\rangle$, where $|S^c\rangle= H^{\otimes m}|0^m\rangle=\frac{1}{\sqrt{n}}\sum_{d=1}^{n}|d\rangle \quad (m^2 = n)$, $|\psi\rangle= \mathcal{U}|0^n\rangle=\sum_{x=0}^{N-1}a_x|\vec{x}\rangle$, $H$ is the Hadamard gate, and $\mathcal{U}$ is an arbitrary unitary quantum gate.

\item [2.] Apply a product of arbitrary local operations $V_{1}\otimes V_{2}\otimes \cdots \otimes V_{n}$ on the initial state $|\psi\rangle$, where $V_{j}$ is chosen from the three single-qubit Pauli gates (X, Y, and Z) acting on the $j$th qubit. The resulting state is $|S^c\rangle\otimes(V_{1}\otimes V_{2}\otimes \cdots \otimes V_{n})|\psi\rangle$.

\item [3.] Perform a Hadamard gate $H$ to each qubit in the register. The resulting state is $|S^c\rangle\otimes H^{\otimes{n}}(V_{1}\otimes V_{2}\otimes \cdots \otimes V_{n})|\psi\rangle$.

\item [4.] Repeat $\tau$ times of the perturbed evolution operator $\mathcal{V}=\mathcal{S}\mathcal{C}$. 

\item [5.] Measure the state of the register. 
\end{itemize}
\label{alg:SKW-3}
\end{algorithm}

\begin{theorem}\label{thm3}
For any given initial state $|\psi\rangle$, the success probability of the SKW-3 algorithm, averaging over all $N$ possible target states, is upper bounded after applying $O(\sqrt{N})$ iterations by
\begin{equation}\label{max-3}
P_{\text{max-3}}=\frac{1-C_f^2(|\psi\rangle)}{2},
\end{equation}
where $C_f(|\psi\rangle)$ is the coherence defined in Eq.(\ref{coh}). 
\end{theorem}

Since $0 \leqslant C_f(|\psi\rangle) \leqslant 1$, it follows that $0 \leqslant P_{\text{max-3}} \leqslant \frac{1}{2}$. Theorem 3 shows that $P_{\text{max-3}}$ depends on the coherence measure $C_f(|\psi\rangle)$ of the initial register state $|\psi\rangle$. If the initial state is a pure incoherent state ($C_f(|\psi\rangle) = 0$), then $P_{\text{max-3}}$ equals $1/2$. If the initial state is a coherent state, the success probability of the SKW-3 algorithm will be less than $1/2$. If the initial state is a maximally coherent state ($C_f(|\psi\rangle) = 1$), the success probability becomes zero, indicating no chance of success. Similar results may be derived for a similarly modified OSKW algorithm, with success probability $P_{\text{max}}=1-C_f^2(|\psi\rangle)$.

Theorem \ref{thm3} reveals a fundamental trade-off in the SKW-3 algorithm: higher quantum coherence in the initial state suppresses the success probability, while lower coherence enhances it. This challenges the conventional view that coherence always benefits quantum computation, suggesting that controlled decoherence may, in some cases, be advantageous. The result provides new insights into quantum algorithm design, emphasizing coherence optimization as a key factor. 

\noindent\textbf{Relationships among the three modified algorithms --}\label{rel}
By generalizing the SKW framework, the SKW-1 algorithm replaces the Hadamard gate with an arbitrary unitary $\mathcal{U}$ by preparing the initial state $|\psi\rangle = \sum_{x=0}^{N-1} a_x |\vec{x}\rangle$. Notably, its success probability depends solely on the coherence fraction of the initial state, rather than its entanglement or coherence. Building upon the SKW-1 algorithm, we introduce the other two modifications: the SKW-2 and SKW-3 algorithms. These refinements extend the framework, enabling a deeper analysis of the roles of entanglement and coherence in algorithmic performance. Now, we examine the relationships among the three algorithms in more detail.

From the analysis of the SKW-1 algorithm, we observe that the maximal coherent state yields the highest success probability. However, in the SKW-3 algorithm, a pure incoherent state gives rise to the highest success probability. This discrepancy arises from the application of the Hadamard gate $H$ to each qubit in the node space register within the SKW-3 algorithm. This operation enhances the coherence of the initially incoherent state while simultaneously reducing the coherence of an initially coherent state. The resulting state after this transformation can be considered equivalent to the initial state of the SKW-1 algorithm, which explains the contrasting outcomes observed in the two cases. 

By integrating the SKW-2 and SKW-3 algorithms, it becomes evident that the SKW-3 algorithm is simply a specialized case of the SKW-2 algorithm. It is straightforward to observe that the combination of a single-qubit Pauli gate and a Hadamard gate retains its character as a local unitary operator. Furthermore, within the computational basis, all incoherent states are inherently separable. These two observations collectively reinforce the conclusion that the SKW-3 algorithm is, in fact, a specific instance of the SKW-2 algorithm. Since the set of incoherent states $\mathcal{I}$ is a proper subset of the set of separable states $\mathsf{S}$, the latter encompasses a larger space of possible choices when minimizing. As a result, we obtain the inequality 
$\min_{\sigma \in \mathsf{S}} \sqrt{1 - F(\sigma, |\psi\rangle)} \leq \min_{\delta \in \mathcal{I}} \sqrt{1 - F(\delta, |\psi\rangle)}$.
This leads to the conclusion that
$E_g(|\psi\rangle) \leq C_f(|\psi\rangle)$, 
and $P_{\text{max-3}} \leq P_{\text{max-2}}$.
This implies that the success probability of applying an arbitrary local unitary operation is higher in the SKW-2 algorithm.

\begin{figure}[h]
\centering
\includegraphics[width=2.7in]{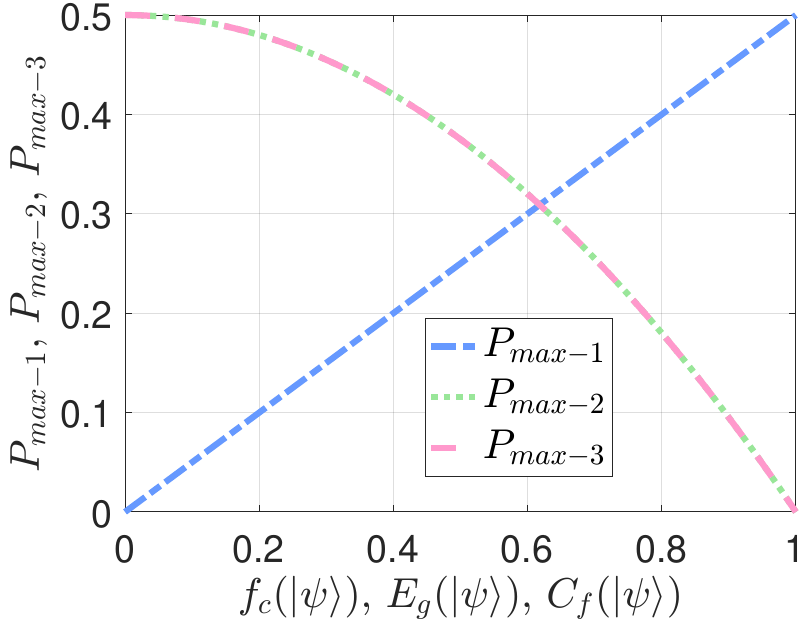}
\caption{\textbf{The relationship among the success probabilities of the three modified SKW algorithms and the properties of the initial states.} With respect to  Eq.~(\ref{max-1}), (\ref{max-2}) and (\ref{max-3}), the blue, green and pink lines represent the relationships between the success probabilities of the SKW-1, SKW-2 and SKW-3 algorithms and the coherence fraction, entanglement and coherence of the initial state, respectively.}
\label{fig:SKW-4}
\vspace{-4mm}
\end{figure}

For clarity, we provide a diagrammatic sketch of the success probabilities of the three modified SKW algorithms in Fig.~\ref{fig:SKW-4}. The success probability of SKW-1 algorithm increases with the coherence fraction of the initial state, while SKW-2 and SKW-3 algorithms show a decrease in success probability as the entanglement and coherence of the initial state increase. Our results highlight the distinct sensitivities of the algorithms to these properties of the initial state.

Our analysis shows that the performance of the SKW algorithm is mainly determined by the coherence fraction of the initial state, rather than its coherence or entanglement. This provides practical guidance for quantum search algorithm design. For instance, in SKW-1 algorithm, where no quantum operation is applied before the perturbed evolution, a higher coherence fraction improves success probability. In SKW-2 algorithm, applying a unitary operation before iteration, lower entanglement enhances performance. Similarly, in SKW-3 algorithm, where a different unitary is applied, lower quantum coherence appears more favorable. These findings suggest that the effectiveness of different quantum properties depends not only on their type but also on the structure of the algorithm, particularly on whether and how the initial state is transformed before the main quantum walk evolution.

Finally, in quantum search algorithms, query complexity serves as the central metric for evaluating algorithmic efficiency. For SKW-1, SKW-2, and SKW-3, it remains $O(\sqrt{N})$, consistent with the original SKW algorithm. However, the time complexity may differ. The original SKW algorithm initializes the system using a single layer of Hadamard gates, whereas the modified versions rely on an arbitrary $n$-qubit unitary $\mathcal{U}$, whose implementation typically requires a quantum circuit of exponential depth in $n$.

\noindent\textbf{Discussions --}\label{dis}
We have investigated how the coherence fraction, entanglement, and coherence affect the success probabilities in three modified quantum random-walk search algorithms. Our analysis reveals that the performance of the SKW algorithm is more closely related to the coherence fraction rather than the entanglement and coherence. This finding highlights the nuanced role of entanglement and coherence in the initial state, which may reduce the algorithm's effectiveness under certain conditions. These results offer a clearer view of the factors driving quantum computational advantage and suggest new strategies for optimizing quantum search algorithms. 

The three algorithms proposed in this work are modified versions of the SKW algorithm, a quantum search framework based on quantum random walks. It offers a quantum speedup similar to Grover’s algorithm for unstructured database search. This efficiency makes it a promising candidate for artificial intelligence applications, particularly in machine learning~\cite{jordan2015machine,das2019machine}. The SKW algorithm may enhance quantum support vector machines~\cite{PhysRevLett.113.130503} through faster feature selection, improve hyperparameter tuning in quantum neural networks~\cite{doi:10.34133/research.0134}, accelerate convergence in deep learning~\cite{lecun2015deep}, and boost exploration efficiency in quantum reinforcement learning~\cite{li2020quantum}. As quantum computing advances, the SKW algorithm is expected to offer a more efficient computational framework for artificial intelligence.

Our findings not only advance the theoretical understanding of quantum walk based algorithms, but also offer actionable guidelines for developing practical quantum algorithms. Furthermore, investigating quantum algorithms whose performance relies on the interplay of coherence, entanglement, and other quantum resources presents a promising research direction. The demonstrated connections between quantum resources and computational performance provide a valuable framework for engineering quantum advantages in various applications, particularly in quantum-enhanced machine learning and artificial intelligence where search problems are ubiquitous. As quantum technologies advance, these insights may prove essential for unlocking quantum advantages in solving complex problems that remain intractable for classical computers. However, the increased circuit depth and control precision required by the modified algorithms significantly raise experimental overhead and sensitivity to noise, presenting new challenges beyond those encountered in previous SKW implementations~\cite{PhysRevA.81.022308}. Addressing these obstacles will be a key direction for future work in circuit optimization and error mitigation.

\section*{Acknowledgements}
This work was supported by the Fundamental Research Funds for the Central Universities, the National Natural Science Foundation of China (12371132, 12075159, 12171044, 12071179, 12405006), and the specific research fund of the Innovation Platform for Academicians of Hainan Province.

\bibliographystyle{iopart-num}
\bibliography{SKW}

\clearpage\clearpage
\newpage

\maketitle
\onecolumngrid

\begin{center}\large \textbf{Decoding Quantum Search Advantage: The Critical Role of State Properties in Random Walks} \\
\textbf{--- Supplementary Material ---}\\
\end{center}

\section{Proof of Theorem 1}\label{I}
\begin{pf}
We begin by preparing the quantum system in the state
\begin{equation}
    |\Phi\rangle=|S^c\rangle \otimes |\psi\rangle,
\end{equation}
where the direction space state $|S^c\rangle$ and the node space initial state $|\psi\rangle$ are constructed as follows.

The direction state is defined by
\begin{equation}
    |S^c\rangle := \frac{1}{\sqrt{n}}\sum_{d=1}^{n}|d\rangle,
\end{equation}
which represents an equal superposition over all directions on the hypercube. This state can be efficiently prepared by applying Hadamard gates to an input state $|0^m\rangle$, where $m^2=n$, thus requiring $m$-qubit Hadamard operations on the direction space.

The initial node state is taken to be an arbitrary pure state
\begin{equation}
    |\psi\rangle=\sum_{x=0}^{N-1}a_x|\vec{x}\rangle,
\end{equation}
where the amplitudes $\{a_x\}$ satisfy $|\sum_{x=0}^{N-1}a_x|=1$. This state can be accomplished efficiently on the node space by applying an arbitrary unitary quantum gate $\mathcal{U}$ to the $|\vec{0^n}\rangle$ state, i.e.,
\begin{equation}
    |\psi\rangle=\mathcal{U}|\vec{0^n}\rangle.
\end{equation}

Having initialized the system to $|\Phi\rangle$, we proceed with the perturbed quantum walk evolution defined by the operator $\mathcal{V}$, iterated for $O(\sqrt{N})$ steps. The goal is to evaluate the maximal success probability averaged over all target states of finding the system in the marked state after the quantum walk evolution.

Let the target state corresponding to the marked vertex $\vec{x}_{\mathrm{tg}}\in\{0, 1\}^{n}$ be defined as
\begin{equation}
|\Gamma\rangle:=|S^c\rangle \otimes |\vec{x}_{\mathrm{tg}}\rangle.
\end{equation}
 
The maximal success probability $P_{\text{max-1}}$, averaging over all $N=2^{n}$ possible target states by using the uniform distribution, is thus given by
\begin{equation}
\begin{aligned}\label{max-1-1}
P_{\text{max-1}}&=\frac{1}{N}\sum_{|\Gamma\rangle}|\langle \Gamma|\mathcal{V}^{\tau} (|S^c\rangle \otimes \mathcal{U}|\vec{0^{n}}\rangle)|^{2}\\
&=\frac{1}{N}\sum_{|\Gamma\rangle}|\langle \Gamma|\mathcal{V}^{\tau} (|S^c\rangle \otimes|\psi \rangle)|^{2}\\
&=\frac{1}{N}\sum_{|\Gamma\rangle}|\langle \Gamma|\mathcal{V}^{\tau} |\Phi\rangle|^{2}.
\end{aligned}
\end{equation}

To analyze Eq.(\ref{max-1-1}), we consider the action of the perturbed evolution operator $\mathcal{V}$ on the equal superposition state 
\begin{equation}
|S^s\rangle=|\eta\rangle, \quad \text{where} \quad
|\eta\rangle := \frac{1}{\sqrt{N}} \sum_{x=0}^{N-1} |\vec{x}\rangle,
\end{equation} 
which is used in the original SKW algorithm. 

According to the analysis presented in the original SKW algorithm~\cite{PhysRevA.67.052307}, it is known that applying the perturbed evolution operator $\mathcal{V}$ for an optimal number of iterations 
\begin{equation}
\tau_{\text{opt}}=\frac{\pi}{2}\sqrt{2^{n-1}}
\end{equation} 
to the system state 
\begin{equation}
|\Upsilon\rangle=|S^c\rangle \otimes |\eta\rangle,
\end{equation} 
we get 
\begin{equation}
\mathcal{V}^{\tau_{\text{opt}}}|\Upsilon\rangle=\frac{1}{\sqrt{2}}|\Gamma\rangle+O(\frac{1}{\sqrt{N}}).
\end{equation} 

The second term is a small correction due to the fact that the SKW algorithm does not yield a solution with probability $1/2$, but rather with probability 
\begin{equation}
\frac{1}{2}-O(\frac{1}{\sqrt{N}}).  
\end{equation}

Multiplying this equation by $(\mathcal{V}^{\tau_{\text{opt}}})^{\dagger}$ and then taking the Hermitian conjugate, we have 
\begin{equation}
\langle \Gamma|\mathcal{V}^{\tau_{\text{opt}}}=\frac{1}{\sqrt{2}} \langle\Upsilon|+O(\frac{1}{\sqrt{N}}). 
\end{equation} 

Substituting this into Eq.(\ref{max-1-1}), we obtain
\begin{equation}\label{max-1-2}
\begin{aligned}
P_{\text{max-1}}
&=\frac{1}{2N}\sum_{|\Gamma\rangle}|\langle\Upsilon|\Phi\rangle|^{2}+O(\frac{1}{\sqrt{N}})\\
&=\frac{1}{2N}\sum_{|\Gamma\rangle}|(\langle s^{C}| \otimes \langle \eta|)(|s^{C}\rangle \otimes |\psi \rangle) |^{2}+O(\frac{1}{\sqrt{N}}).
\end{aligned}
\end{equation}

Since $|\Upsilon\rangle = |S^c\rangle \otimes |\eta\rangle$ and $|\Phi\rangle = |S^c\rangle \otimes |\psi\rangle$, we compute the inner product
\begin{equation}
\langle \Upsilon | \Phi\rangle = \langle S^c | S^c\rangle \cdot \langle \eta | \psi\rangle = \langle \eta | \psi\rangle,
\end{equation}
implying that the inner product is independent of the specific target state. 

Thus we have
\begin{equation}\label{max-1-3}
   P_{\text{max-1}}(|\psi\rangle)
   =\frac{1}{2}|\langle\eta| \psi \rangle |^{2}+O(\frac{1}{\sqrt{N}}). 
\end{equation}

We define the coherence fraction of the initial state $|\psi\rangle$ as
\begin{equation}
f_c(|\psi\rangle) := |\langle \eta | \psi\rangle|^2,
\end{equation}
which quantifies the overlap of the initial state with the equal superposition state (maximal coherence state) $|\eta\rangle$.

Therefore, we obtain the desired result
\begin{equation}
   P_{\text{max-1}}(|\psi\rangle)
   =\frac{1}{2}f_{c}(|\psi\rangle)+O(\frac{1}{\sqrt{N}}),  
\end{equation}
which completes the proof of Theorem 1. 
\hfill $\square$
\end{pf}

\section{The success probability of SKW-1 algorithm for mixed initial states}\label{II}
We now extend the result of Theorem 1 to the case where the initial state in the node space is a general $n$-qubit mixed state. Let the initial state of the quantum register be
\begin{equation}
\rho=\sum_{\mu}p_{\mu}|\psi_{\mu}\rangle\langle\psi_{\mu}|,
\end{equation}
where each $|\psi_\mu\rangle$ is a pure state in the node space, $p_\mu \geq 0$, and $\sum_\mu p_\mu = 1$. Without loss of generality, each $|\psi_\mu\rangle$ may be written in the computational basis as
\begin{equation}
|\psi_\mu\rangle = \sum_{i=0}^{N-1} a_{\mu i} |i\rangle,
\quad \text{with} \quad \sum_i |a_{\mu i}|^2 = 1.
\end{equation}
The full state of the system is therefore given by
\begin{equation}
\rho_{\mathrm{in}} = |S^c\rangle \langle S^c| \otimes \rho.
\end{equation}

Let $|\Gamma\rangle = |S^c\rangle \otimes |\vec{x}_{\mathrm{tg}}\rangle$ represent the target state corresponding to a marked vertex $\vec{x}_{\mathrm{tg}} \in \{0,1\}^n$. The optimal number of iterations of the perturbed quantum walk operator $\mathcal{V}$ is $\tau_{\mathrm{opt}} = \frac{\pi}{2} \sqrt{2^{n-1}}$. Then the average success probability after applying $\mathcal{V}^{\tau_{\mathrm{opt}}}$ is given by
\begin{equation}
P_{\text{max-1}}(\rho) = \frac{1}{N} \sum_{\vec{x}_{\mathrm{tg}}} \operatorname{Tr}\left[ \mathcal{V}^{\tau_{\mathrm{opt}}} \rho_{\mathrm{in}} (\mathcal{V}^{\tau_{\mathrm{opt}}})^\dagger \cdot |\Gamma\rangle\langle\Gamma| \right].
\end{equation}

As shown previously in the pure-state case, we may approximate
\begin{equation}
\langle \Gamma | \mathcal{V}^{\tau_{\mathrm{opt}}} = \frac{1}{\sqrt{2}} \langle \Upsilon | + O(\frac{1}{\sqrt{N}}),
\end{equation}
where $|\Upsilon\rangle = |S^c\rangle \otimes |\eta\rangle$, with $|\eta\rangle = \frac{1}{\sqrt{N}} \sum_{i=0}^{N-1} |i\rangle$ as before. Substituting, we obtain

\begin{equation}
\begin{aligned}
P_{\text{max-1}}(\rho)
&= \frac{1}{N} \sum_{\vec{x}_{\mathrm{tg}}} \left| \langle \Gamma | \mathcal{V}^{\tau_{\mathrm{opt}}} \rho_{\mathrm{in}} (\mathcal{V}^{\tau_{\mathrm{opt}}})^\dagger | \Gamma \rangle \right| \\
&= \frac{1}{N} \sum_{\vec{x}_{\mathrm{tg}}} ( \frac{1}{\sqrt{2}} \langle \Upsilon | + O(\frac{1}{\sqrt{N}})) \rho_{\mathrm{in}} ( \frac{1}{\sqrt{2}} | \Upsilon \rangle + O(\frac{1}{\sqrt{N}})) \\
&= \frac{1}{2} \langle \Upsilon | \rho_{\mathrm{in}} | \Upsilon \rangle + O(\frac{1}{\sqrt{N}}).
\end{aligned}
\end{equation}

Using the tensor product structure of $\rho_{\mathrm{in}}$, we compute
\begin{equation}
\langle \Upsilon | \rho_{\mathrm{in}} | \Upsilon \rangle = \langle S^c | S^c \rangle \cdot \langle \eta | \rho | \eta \rangle = \langle \eta | \rho | \eta \rangle.
\end{equation}

Thus, the average success probability becomes
\begin{equation}
P_{\text{max-1}}(\rho) = \frac{1}{2} \langle \eta | \rho | \eta \rangle + O(\frac{1}{\sqrt{N}}).
\end{equation}

We define the coherence fraction of the mixed state $\rho$ as
\begin{equation}
f_c(\rho) := \langle \eta | \rho | \eta \rangle,
\end{equation}
which quantifies the overlap of the initial state with the equal superposition state (maximal coherence state). Therefore,

\begin{equation}
P_{\text{max-1}}(\rho) = \frac{1}{2} f_c(\rho) + O(\frac{1}{\sqrt{N}}),
\end{equation}
which completes the generalization of Theorem 1 to mixed initial states. \hfill $\square$
\section{The success probability of the optimized SKW-1 algorithm}\label{III}
Based on the optimized version of the SKW algorithm (OSKW)~\cite{PhysRevA.79.012325} and the SKW-1 algorithm, we consider an optimized version of the SKW-1 algorithm, denoted as OSKW-1 algorithm, which operates on an $(n+1)$-dimensional hypercube and restricts the evolution to the subspace of even-parity vertices. The system is initialized in the state
\begin{equation}
|\Upsilon_{\mathrm{opt}}\rangle = |S^c_{\mathrm{opt}}\rangle \otimes |\psi\rangle = |S^c_{\mathrm{opt}}\rangle \otimes \mathcal{U}|\vec{0}^{n+1}\rangle,
\end{equation}
where $|S^c_{\mathrm{opt}}\rangle = \frac{1}{\sqrt{n+1}} \sum_{d=1}^{n+1} |d\rangle$ is the equal superposition over the direction space, and $|\psi\rangle \in \mathbb{C}^{2^{n+1}}$ is an arbitrary pure state in the node space, prepared via an arbitrary unitary gate $\mathcal{U}$. 

A projection operator $\mathcal{P}^e$ is applied to ensure that the position state lies in the even-parity subspace, i.e.,
\begin{equation}
\mathcal{P}^e := \sum_{\substack{x \in \{0,1\}^{n+1}\\ |\vec{x}| \equiv 0 \ (\mathrm{mod}\ 2)}} |\vec{x}\rangle\langle\vec{x}|,
\end{equation}
and the normalized projected state is defined as $|\psi^e\rangle := \mathcal{P}^e |\psi\rangle / \|\mathcal{P}^e |\psi\rangle\|$. 

The perturbed evolution operator is given by 

\begin{equation}
\mathcal{V}_{\mathrm{opt}} = \mathcal{S} (\mathcal{C}_0 \otimes \mathbf{I}) \mathcal{S} \mathcal{C},
\end{equation}
and the target state is 
\begin{equation}
|\Gamma_{\mathrm{opt}}\rangle = |S^c_{\mathrm{opt}}\rangle \otimes |\vec{x}_{\mathrm{tg}}\rangle,
\end{equation} 
where $|\vec{x}_{\mathrm{tg}}| \equiv 0 \ (\mathrm{mod}\ 2)$. 

The average success probability over all target vertices after applying $\tau = \tau_{\mathrm{opt}} \sim O(\sqrt{N})$ steps is given by
\begin{equation}
P_{\text{max-1}}^{\text{opt}} = \frac{1}{N} \sum_{\vec{x}_{\mathrm{tg}}} \left| \left\langle \Gamma_{\mathrm{opt}} \middle| \mathcal{V}_{\mathrm{opt}}^{\tau} \mathcal{P}^e |\Upsilon_{\mathrm{opt}}\rangle \right\rangle \right|^2.
\end{equation}

Following the OSKW algorithm, it is known that 
\begin{equation}
\mathcal{V}_{\mathrm{opt}}^\tau ( |S^c_{\mathrm{opt}}\rangle \otimes |\eta^e\rangle) = |\Gamma_{\mathrm{opt}}\rangle + O(\frac{1}{\sqrt{N}}),
\end{equation}
where $|\eta^e\rangle := \frac{1}{\sqrt{N/2}} \sum_{|\vec{x}| \equiv 0 \ (\mathrm{mod}\ 2)} |\vec{x}\rangle $ is the equal superposition over all even-parity vertices. 

Then applying $\tau_{\text{opt}} = \frac{\pi}{2\sqrt{2}} \sqrt{N}$ steps yields
\begin{equation}
\mathcal{V}_{\mathrm{opt}}^{\tau_{\text{opt}}} |\Upsilon_{\eta^e}\rangle = |\Gamma_{\mathrm{opt}}\rangle + O\left( \frac{1}{\sqrt{N}} \right).
\end{equation}

Taking the Hermitian conjugate gives 
\begin{equation}
\langle \Gamma_{\mathrm{opt}} | \mathcal{V}_{\mathrm{opt}}^{\tau_{\text{opt}}} = \langle S^c_{\mathrm{opt}} | \otimes \langle \eta^e | + O(\frac{1}{\sqrt{N}}).
\end{equation}

Substituting into the expression for success probability $P_{\text{max-1}}^{\text{opt}}$, and using $|\Upsilon_{\mathrm{opt}}\rangle = |S^c_{\mathrm{opt}}\rangle \otimes |\psi\rangle$, we obtain
\begin{equation}
\begin{aligned}
P_{\text{max-1}}^{\text{opt}} 
&= \frac{1}{N} \sum_{\vec{x}_{\mathrm{tg}}} \left| \langle S^c_{\mathrm{opt}} | \otimes \langle \eta^e | \mathcal{P}^e |\Upsilon_{\mathrm{opt}}\rangle \right|^2 + O\left( \frac{1}{\sqrt{N}} \right) \\
&= \left| \langle \eta^e | \mathcal{P}^e |\psi\rangle \right|^2 + O\left( \frac{1}{\sqrt{N}} \right).
\end{aligned}
\end{equation}

In the case where the original state $|\psi\rangle$ lies entirely within the even subspace, i.e., $\mathcal{P}^e |\psi\rangle = |\psi\rangle$, this simplifies to
\begin{equation}
P_{\text{max-1}}^{\text{opt}} = \left| \langle \eta^e | \psi \rangle \right|^2 + O(\frac{1}{\sqrt{N}}),
\end{equation}
which we identify as the coherence fraction of $|\psi\rangle$ with respect to the equal superposition over the even-parity subspace. 

Thus, the maximal success probability of the OSKW-1 algorithm satisfies
\begin{equation}
P_{\text{max-1}}^{\text{opt}}(|\psi\rangle) = f_c(|\psi\rangle) + O(\frac{1}{\sqrt{N}}),
\end{equation}
where $f_c(|\psi\rangle) := |\langle \eta^e | \psi \rangle|^2$. 
\hfill$\square$

\section{Proof of Theorem 2}\label{IV}
\begin{pf}
We consider the SKW-2 algorithm, where the initial node state $|\psi\rangle$ is transformed by a layer of local single-qubit unitaries $U_1 \otimes U_2 \otimes \cdots \otimes U_n$, applied before the quantum walk iterations. As in the original SKW algorithm, we denote the target state as
\begin{equation}
  |\Gamma\rangle = |S^c\rangle \otimes |\vec{x}_{\mathrm{tg}}\rangle, 
\end{equation}
where $|S^c\rangle = \frac{1}{\sqrt{n}} \sum_{d=1}^{n} |d\rangle$ is the equal superposition state, and $\vec{x}_{\mathrm{tg}} \in \{0,1\}^n$ is the marked vertex.

Averaging over all $N = 2^n$ possible target states using a uniform distribution, the maximal success probability of the SKW-2 algorithm can be written as
\begin{equation}\label{max-2-1}
P_{\text{max-2}}=\max_{U_{1}, U_{2}, \cdots, U_{n}}\frac{1}{N}\sum_{|\Gamma\rangle}|\langle \Gamma|\mathcal{V}^{\tau_{\mathrm{opt}}} [|S^c\rangle \otimes (U_{1}\otimes U_{2}\otimes \cdots \otimes U_{n})|\psi\rangle]|^{2},
\end{equation}
where $\mathcal{V}$ is the perturbed evolution operator and $\tau_{\mathrm{opt}} \sim O(\sqrt{N})$ is the optimal number of iterations.

According to the proof of Theorem 1, we know that after $\tau_{\mathrm{opt}}$ steps,
\begin{equation}
   \langle \Gamma | \mathcal{V}^{\tau_{\text{opt}}} = \frac{1}{\sqrt{2}} \left( \langle S^c| \otimes \langle \eta| \right) + O\left( \frac{1}{\sqrt{N}} \right),
\end{equation}
where $|\eta\rangle := \frac{1}{\sqrt{N}} \sum_{x=0}^{N-1} |\vec{x}\rangle$ is the equal superposition over the node space. 

Substituting into Eq.(\ref{max-2-1}), we obtain
\begin{equation}\label{max-2-2}
P_{\text{max-2}}=\max_{U_{1}, U_{2}, \cdots, U_{n}}\frac{1}{2N}\sum_{|\Gamma\rangle}|\langle \eta|U_{1}\otimes U_{2}\otimes \cdots \otimes U_{n}|\psi\rangle|^{2}+O(\frac{1}{\sqrt{N}}).
\end{equation}

Since $|\eta\rangle$ is a product state, applying $U_1^\dagger \otimes U_{2}^\dagger \otimes \cdots \otimes U_n^\dagger$ to it yields another product state. Let
\begin{equation}
|u_1, u_2, \dots, u_n\rangle := \left( U_1^\dagger \otimes U_{2}^\dagger \otimes \cdots \otimes U_n^\dagger \right) |\eta\rangle,
\end{equation}
where the maximization now runs over all product states $|u_{1}, u_{2}, \cdots, u_{n}\rangle=|u_{1} \otimes u_{2}\otimes \cdots \otimes u_{n}\rangle$ of the $n$ qubits. The local unitary rotations $U_j$ maps $|u_j\rangle$ to $|+\rangle=\frac{1}{\sqrt{2}}(|0\rangle+|1\rangle)$. If the input state is a product state $|u_{1} \otimes u_{2}\otimes \cdots \otimes u_{n}\rangle$, then $P_{max}$ equals to $1/2$. If the input state is not a product state, the success probability of the SKW-2 algorithm is less than $1/2$.

Then Eq.(\ref{max-2-2}) becomes
\begin{equation} \label{max-2-3}
P_{\text{max-2}}=\max_{|u_{1}, u_{2}, \cdots, u_{n}\rangle}\frac{1}{2}|\langle u_{1}, u_{2}, \cdots, u_{n}|\psi\rangle|^2+O(\frac{1}{\sqrt{N}}),
\end{equation}
where the maximization is taken over all $n$-qubit product states.

Now, let us recall the definition of the Groverian entanglement measure of a state $|\psi\rangle$~\cite{PhysRevA.65.062312}
\begin{equation}
E_g(|\psi\rangle)\equiv \min _{\sigma \in \mathsf{S}} \sqrt{1-F(\sigma, |\psi\rangle)}= \min_{|u_1, u_2, \dots, u_n\rangle \in \mathsf{S}} \sqrt{1 - |\langle u_1, u_2, \dots, u_n | \psi\rangle|^2},
\end{equation}
where $\mathsf{S}$ is the set of all separable states.

Squaring both sides and rearranging, we get
\begin{equation}
\max_{|u_1, u_2, \dots, u_n\rangle} |\langle u_1, u_2, \dots, u_n | \psi\rangle|^2 = 1 - E_g^2(|\psi\rangle).
\end{equation}

Substituting into Eq.(\ref{max-2-3}), and neglecting the term $O(1/\sqrt{N})$, we have
\begin{equation}
P_{\text{max-2}} = \frac{1 - E_g^2(|\psi\rangle)}{2}.
\end{equation}
This completes the proof of Theorem 2.

\hfill $\square$
\end{pf}

\bigskip
\section{Proof of Theorem 3}\label{V}
\begin{pf}
We now turn to the SKW-3 algorithm, which differs from SKW-1 and SKW-2 in the quantum operation applied to the initial state. Specifically, it applies a layer of arbitrary local unitaries $V_{1}\otimes V_{2}\otimes \cdots \otimes V_{n}$, followed by Hadamard gates $H^{\otimes n}$, prior to the quantum walk evolution. The local unitary $V_{j}$ is chosen from the three single-qubit Pauli gates (X, Y, and Z) acting on the $j$th qubit.

The complete initial state for the algorithm is then prepared as
\begin{equation}
|\Phi\rangle = |S^c\rangle \otimes H^{\otimes n}(V_{1}\otimes V_{2}\otimes \cdots \otimes V_{n}) |\psi\rangle,
\end{equation}
where $|S^c\rangle = \frac{1}{\sqrt{n}} \sum_{d=1}^{n} |d\rangle$ is the equal superposition in the coin space, and $|\psi\rangle$ is an arbitrary initial state on the node space.

We define the target state as
\begin{equation}
|\Gamma\rangle = |S^c\rangle \otimes |\vec{x}_{\mathrm{tg}}\rangle,
\end{equation}
where $\vec{x}_{\mathrm{tg}} \in \{0,1\}^n$ denotes the marked vertex. 

Averaging uniformly over all $N = 2^n$ possible target vertices, the maximal success probability of the SKW-3 algorithm is given by
\begin{equation}\label{max-3-1}
P_{\text{max-3}}=\max_{V_{1}, V_{2}, \cdots, V_{n}}\frac{1}{N}\sum_{|\Gamma\rangle}|\langle \Gamma|\mathcal{V}^{\tau_{\mathrm{opt}}} [|S^c\rangle \otimes H^{\otimes{n}}(V_{1}\otimes V_{2}\otimes \cdots \otimes V_{n})|\psi\rangle]|^{2},
\end{equation}
where $\mathcal{V}$ is the perturbed quantum walk operator, and $\tau_{\mathrm{opt}} = \frac{\pi}{2\sqrt{2}} \sqrt{N}$ is the optimal number of iterations.

Following the analysis in Theorem 1, we approximate
\begin{equation}
\langle \Gamma|\mathcal{V}^{\tau_{\mathrm{opt}}}=\frac{1}{\sqrt{2}} (\langle S^c| \otimes \langle \eta|)+O(1/\sqrt{N}),
\end{equation}
where $|\eta\rangle = \frac{1}{\sqrt{N}} \sum_{x=0}^{N-1} |\vec{x}\rangle = |+\rangle^{\otimes n}$ is the equal superposition in the node space.

Substituting it into Eq.(\ref{max-3-1}), we get
\begin{equation}\label{max-3-2}
P_{\text{max-3}}=\max_{V_{1}, V_{2}, \cdots, V_{n}}\frac{1}{2N}\sum_{|\Gamma\rangle}|\langle 0^{n}|(V_{1}\otimes V_{2}\otimes \cdots \otimes V_{n})|\psi\rangle|^{2}+O(\frac{1}{\sqrt{N}}).
\end{equation}

Note that $|\eta\rangle=|+\rangle^{\otimes{n}}$. We have 
\begin{equation}
H^{\otimes{n}}|\eta\rangle=|0^n\rangle,
\end{equation}
and 
\begin{equation}
(V_{1}\otimes V_{2}\otimes \cdots \otimes V_{n})|0^n\rangle=|i\rangle, \quad (i=0, 1, \cdots, N-1),
\end{equation}
where the maximization now runs over all basis states $|i\rangle$ of the $n$ qubits. Since local unitaries map the computational basis vectors onto other orthonormal basis vectors, the maximization is equivalent to a maximization over computational basis states $|i\rangle$. Therefore, the optimization in Eq.({\ref{max-3-2}}) can be equivalently expressed as an optimization over basis states,
\begin{equation}\label{max-3-3}
    P_{\text{max-3}}=\max_{|i\rangle}\frac{1}{2}|\langle i|\psi\rangle|^2+O(\frac{1}{\sqrt{N}}),
\end{equation}
where the maximization now runs over all basis states $|i\rangle$ of the $n$ qubits. 

The maximization in Eq.({\ref{max-3-3}}) can be transformed into a maximization over the set $\mathcal{I}$ of all incoherent states,
\begin{equation}\label{max-3-4}
    P_{\text{max-3}}=\max_{\delta\in \mathcal{I}}\frac{1}{2}\langle\psi| \delta|\psi\rangle+O(\frac{1}{\sqrt{N}}),
\end{equation}
which can be further rewritten as
\begin{equation}\label{max-3-5}
    P_{\text{max-3}}=\max_{\delta\in \mathcal{I}}\frac{1}{2}F(\delta, |\psi\rangle)+O(\frac{1}{\sqrt{N}}).
\end{equation}

Now recall that the  fidelity-based measure of coherence is given by~\cite{liu2017new}
\begin{equation}\label{coh}
C_f(|\psi\rangle) \equiv \min_{\delta \in \mathcal{I}} \sqrt{1 - \langle \psi | \delta | \psi \rangle},
\end{equation}
where $\mathcal{I}$ denotes the set of incoherent (diagonal) states. Using the fact that for pure states, the optimal $\delta \in \mathcal{I}$ is a basis state $|i\rangle\langle i|$, we have
\begin{equation}
\max_{|i\rangle} |\langle i | \psi\rangle|^2 = 1 - C_f^2(|\psi\rangle).
\end{equation}

By substituting Eq.(\ref{coh}) into Eq.(\ref{max-3-5}) and neglecting the term $O(1/\sqrt{N})$, we establish a relationship between the success probability of the SKW-3 algorithm $P_{\text{max-3}}$, and the coherence $C_f(|\psi\rangle)$,  which yields:
\begin{equation}
    P_{\text{max-3}}=\frac{1-C_f^2(|\psi\rangle)}{2}.
\end{equation}
This concludes the proof of Theorem 3. 
\hfill $\square$
\end{pf}

\end{document}